\documentstyle[aps,manuscript]{revtex}

\begin{document}

\title {Generalized strategies in the Minority Game}
\author{M. Hart, P. Jefferies and N.F. Johnson}
\address {Physics Department, Oxford University, Oxford, OX1 3PU, U.K.}
\author{P.M. Hui}
\address {Department of Physics, The Chinese University of Hong Kong, Shatin,
\\ New Territories, Hong Kong}

\maketitle

\begin{abstract} We show analytically how the fluctuations (i.e. standard
deviation $\sigma$) in the Minority Game (MG) can decrease below
the random coin-toss limit if the agents use more general, stochastic
strategies.  This suppression of $\sigma$ results from a cancellation between
the actions of a crowd,  in which agents act collectively and make the same
decision,  and an anticrowd in which agents act collectively by making the
opposite decision to the crowd. 

\end{abstract}
\bigskip

\noindent {\bf Phys. Rev. E (in press)}

\noindent {\em cond-mat/0006141}

\newpage

\baselineskip=10pt

The Minority Game (MG) of Challet and Zhang\cite{challet,challet2,savit}
offers a simple paradigm for complex, adaptive systems. The MG comprises an
odd number
$N$ of agents, each with $s$ strategies and a memory size
$m$, who repeatedly compete to be in the minority. In the basic MG, where
agents always use their highest scoring strategy, the size of the
fluctuations (i.e. standard deviation $\sigma$) falls below the random,
coin-toss limit for large $m$ \cite{savit}. Cavagna {\em et al}
\cite{sherrington} considered an interesting modification of the basic MG,
the `Thermal Minority Game' (TMG), whereby  agents choose between their
strategies using an exponential probability weighting.   
As pointed out by Marsili {\em et al} \cite{marsili}, such a probabilistic strategy
weighting has a long tradition in economics and encodes a particular behavioral model. 
The numerical
simulations of Cavagna {\em et al} demonstrated that at small $m$, where the MG 
$\sigma$ is larger-than-random, the TMG
$\sigma$ could be pushed below the random coin-toss limit just by
altering this relative probability weighting, or equivalently the `temperature' $T$
\cite{sherrington}. This reduction in $\sigma$ for
stochastic strategies seems fairly general: for example, we had reported earlier on a
modified MG in which agents with stochastic strategies also generate a smaller-than-random
$\sigma$ for small $m$ \cite{prl}. 

In this paper, we provide a quantitative theory which explains how the
standard deviation $\sigma$ in the MG can get reduced from larger-than-random to
smaller-than-random if more general, stochastic strategies are used.  In particular, we
show that stochastic strategy rules tend to {\em increase} the cancellation between the
actions of a crowd of like-minded agents \cite{crowd,us}  and its anti-correlated
partner (anticrowd), thereby reducing $\sigma$ below the random coin-toss
limit. 

The MG\cite{challet} comprises an odd number of agents $N$ who choose
repeatedly between option 0 (e.g. buy) and option 1 (e.g. sell). The winners
are those in the minority group, e.g. sellers win if there is an excess of
buyers. The outcome at each timestep represents the winning decision, 0 or 1.
A common bit-string of the $m$ most recent outcomes is made available to the
agents at each timestep \cite{memory}. The agents randomly pick $s$ strategies at the
beginning of the game, with repetitions allowed, from the pool of all
possible strategies.   After each turn, the agent assigns one (virtual) point
to each of his strategies which would have predicted the correct outcome. In
the basic MG, each agent uses the most successful strategy in his possession,
i.e. the one with the  most virtual points. Because of crowd effects
\cite{us}, $\sigma$ is larger-than-random for small $m$. Our task is to
explain quantitatively why the larger-than-random
$\sigma$ in this `crowded' regime (i.e. small $m$) becomes smaller-than-random
when the strategy-picking rule is made stochastic, such as in Ref.
\cite{sherrington}. 

Consider any two strategies $r$ and $r^*$ within the list of $2^{m+1}$
strategies in the reduced strategy space \cite{challet,us}. At any moment in
the game, the strategies can be ranked according to their virtual points,
$r=1,2
\dots 2^{m+1}$ where $r=1$ is the best strategy, $r=2$ is second best, etc.
In the small $m$ regime of interest, the virtual-point strategy
ranking and popularity ranking for strategies are similar \cite{us}. Consider $s=2$ as in
Ref.
\cite{sherrington}. Let
$p(r,r^*|r^*\geq r)$ be the probability that a given agent picks $r$ and
$r^*$, where $r^*\geq r$ (i.e. $r$ is the best, or equal best, among his $s=2$
strategies). In contrast, let
$p(r,r^*|r^*\leq r)$ be the probability that a given agent picks $r$ and
$r^*$, where $r^*\leq r$ (i.e. $r$ is the worst, or equal worst, among his
$s=2$ strategies). Let
$\theta$ be the probability that the agent uses the worst of his $s=2$
strategies, while
$1-\theta$ is the probability that he uses the best.  The probability that
the agent plays
$r$ is given by
\begin{eqnarray} p_r & = & \sum_{r^*=1}^{2^{m+1}} [\ \theta\  p(r,r^*|r^*\leq
r) + \ (1-\theta)
\ p(r,r^*|r^*\geq r)]\nonumber \\ & = & \ (1-\theta)\  p_+(r) + \ \theta\
p_-(r) + 2^{-2(m+1)}\ \theta
\end{eqnarray} where $p_+(r)$ is the probability that the agent has picked
$r$ {\em and} that $r$ is the agent's best (or equal best) strategy; $p_-(r)$
is the probability that the agent has picked
$r$ {\em and} that $r$ is the agent's worst strategy. It is straightforward to
show that
\begin{equation} p_+(r) = \bigg(\bigg[1 - \frac{(r-1)}{2^{m+1}}\bigg]^2 -
\bigg[1-\frac{r}{2^{m+1}}\bigg]^2\bigg)\ \ .
\end{equation} Note that 
$p_+(r) + p_-(r) = p(r)$ where 
\begin{equation} p(r) = 2^{-m} ( 1 - 2^{-(m+2)})
\end{equation} is the probability that the agent holds strategy
$r$ after his $s=2$ picks, with no condition on whether it is best  or
worst.   An expression for $p_-(r)$ follows from Eqs. (2) and (3).   The
basic MG\cite{challet} corresponds to the case $\theta = 0$. 

In the TMG, each agent is
equipped at each timestep with his own (biased) coin characterised by
exponential probability weightings
\cite{sherrington}. An agent then flips this coin at each timestep to decide
which strategy to use
\cite{sherrington}. To relate the present analysis to the TMG in
Ref.\cite{sherrington}, we consider $0\leq\theta \leq 1/2$: 
$\theta=0$ corresponds to `temperature' $T=0$
while
$\theta\rightarrow 1/2$ corresponds to $T\rightarrow \infty$
\cite{tmg} with $\theta= 1/2[1-{\rm tanh} (1/T)]$. Consider the mean number
of agents playing strategy
$r$ which is given by
\begin{equation} n_r = N p_{r} = N\ (1-2\theta)\  p_+(r) + N\ \theta \ p(r) +
2^{-2(m+1)}\ N\ \theta \ \ . 
%2^{-m} (1-2^{-(m+2)})\ \ .
\end{equation} If ${n}_r$ agents use the same strategy $r$, then they will
act as a `crowd', i.e. they will make the same decision. If
${n}_{\bar r}$ agents simultaneously use the strategy
$\bar r$ anticorrelated to $r$,  they will make the opposite
(anticorrelated) decision and will hence act as an `anticrowd' \cite{us}. Averaging over
the stochastics of the strategy-use which is introduced by allowing each agent to flip a
(biased) coin at each timestep, the standard deviation
$\sigma_\theta$ in the  number of agents making a particular decision (say
0)  becomes 
\begin{equation}
\sigma =
\bigg[ \frac{1}{2} \sum_{r=1}^{2^{m+1}} \sum_{r'=1}^{2^{m+1}}
\frac{1}{4}|n_r-n_{r'}|^2 P(r'={\bar r})\bigg]^{\frac{1}{2}}
\end{equation}
where  $P(r'={\bar r})$ is the
probability that any strategy $r'$ is the anti-correlated partner
of strategy $r$ in the list of strategies when ordered in terms of
popularity $\{n_r\}$ \cite{us}. The quantities
$n_r$ and
$n_{r'}$ are
$\theta$-dependent (see Eq. (4)). Substituting Eqs. (3) and (4) for $r$ and $r'$ into Eq.
(5) yields
\begin{equation}
\sigma_\theta = |1-2\theta| \ \sigma_{\theta=0}
\end{equation} where $\sigma_{\theta=0}$ is the standard deviation when
$\theta=0$, i.e. the basic MG. Equation (6) explicitly shows that the standard deviation
$\sigma_\theta$ {\em decreases} as $\theta$ increases (recall $0\leq
\theta
\leq 1/2$): in other words, the standard deviation decreases as agents use
their worst strategy with increasing probability. An increase in $\theta$
leads to a  reduction in the size of the larger crowds using high-scoring
strategies, as well as an increase in the size of the smaller anticrowds using
lower-scoring strategies, hence resulting in a more substantial  cancellation
effect between the crowd and the anticrowd.  As
$\theta$ increases, $\sigma_\theta$ will eventually drop {\em below} the
random coin-toss result at $\theta = \theta_c$ where 
\begin{equation}
\theta_{c} = \frac{1}{2}- \frac{\sqrt{N}}{4}\frac{1}{\sigma_{\theta=0}}\ \ .
\end{equation} 
Elsewhere we presented a quantitative formulation for $\sigma_{\theta=0}$. In
particular, Ref. \cite{us} showed that 
$P(r'={\bar r})$ at small $m$ lay between that of a 
$\delta$-function distribution $\delta_{r',2^{m+1}+1-r}$ peaked at
$r'=2^{m+1}+1-r$, and a flat distribution $[2^{(m+1)}-1]^{-1}$
(N.B. $P(r'={\bar r})=0$ at $r'=r$) \cite{us}. Using these distributions, the
resulting expressions for
$\sigma_\theta$ are
\begin{equation}
\sigma_{{\rm delta}} = (1 - 2\theta) \frac{N}{{\sqrt 3}}
2^{{-(\frac{m}{2}+1)}}
\bigg[ 1 - 2^{-2(m+1)}\bigg] ^{\frac{1}{2}}
\end{equation}
and
\begin{equation}
\sigma_{{\rm flat}} = (1 - 2\theta) \frac{N}{2 \sqrt 3}
[2^{(m+1)} - 1]^{-\frac{1}{2}} 
\bigg[ 1 - 2^{-2(m+1)}\bigg] ^{\frac{1}{2}}
\end{equation}
respectively. For the TMG,  $\theta=(1/2)[1 - {\rm tanh} (1/T)]$ although we emphasize that
the present theory is not limited to the case of `thermal' strategy weightings.

Figure 1 shows a comparison between the theory of Eqs. (8) and (9) and 
numerical simulation for various runs. The theory agrees well in the range
$\theta=0
\rightarrow 0.35$ and, most importantly, provides a quantitative explanation for the
transition in
$\sigma$ from larger-than-random to smaller-than-random as
$\theta$ (and hence $T$) is increased. The numerical data for different runs has a
significant natural spread:
remarkably most of these data points lie in the region of these two
analytic curves. 

Above $\theta=0.35$, the numerical data tend to flatten off while the present theory
predicts a decrease in $\sigma$ as $\theta\rightarrow 0.5$. This is because the present
theory averages out the fluctuations in strategy-use at each time-step (Eq. (4) only
considers the mean number of agents using strategy $r$). Consider $\theta =
0.5$. For a particular configuration of strategies picked at the start of the
game, and at a particular moment in time, the number of agents using each
strategy is typically distributed {\em around} the mean value $n_r=N 2^{-(m+1)}$ given by
Eq. (4) for $\theta=0.5$. The resulting distribution describing the strategy-use is
therefore non-flat. It is these fluctuations about the mean $n_r$ and $n_{\bar r}$ which
give rise to a non-zero
$\sigma$. Our present crowd-anticrowd
theory can be extended to account for the effect of these fluctuations in strategy-use for
$\theta\rightarrow 0.5$ in the following way: All $N$ agents are randomly assigned 2
strategies. To represent a turn in the game, each agent flips a (fair) coin to
decide which of the two strategies is the preferred one. Having generated a list of
the number of agents using each strategy, $\sigma$ is then 
found in the usual way by cancelling off crowds and anticrowds. A time-averaged value for
$\sigma$ is then obtained by averaging over
$100$ independent coin-flip outcomes for the given initial distribution of strategies among
agents. This procedure provides a semi-analytic calculation for the value of
$\sigma$ at $\theta= 0.5$. Inset (a) in Fig. 1 shows the measured numerical
distribution in
$\sigma$ for $\theta = 0.5$, while inset (b) shows the result from the
semi-analytic procedure. The two distributions are in good agreement. It is
also possible to perform a fully analytic calculation of the average
$\sigma_\theta$ in the 
$\theta\rightarrow 0.5$ limit: the
initial random assignment of strategies can be modelled using a random-walk. This
yields an average value of $|n_r -
n_{\bar r}|^2$ (i.e. time-averaged over the fluctuations) given by
$N 2^{-m}  (1 - 2^{-(m+1)})$. Summing over all pairs of
correlated-anticorrelated strategies, i.e. all crowd-anticrowd pairs, yields a
(configuration-average) standard deviation given by 
\begin{equation}
\sigma_{\theta\rightarrow 0.5} = \frac{\sqrt N}{2} \big[ 1 -
2^{-(m+1)}\big]^{\frac{1}{2}}\ \ .
\end{equation}
For $N=101$ as shown in Fig. 1, the analytic theory (Eq. (10)) and semi-analytic theory
both give an average standard deviation $\sigma_{\theta\rightarrow 0.5}=4.7$, while the
numerical simulation yields $4.5$. The agreement is good. We note that all these values lie
{\em below} the random coin-toss limit
${\sqrt N}/2 = 5.0$: hence the theoretical and
numerical results are {\em not} equal to the random coin-toss limit
as
$\theta\rightarrow 0.5$ (i.e. $T\rightarrow\infty$). We have also checked this for
$N=11$: here the analytic value is 1.55, the semi-analytic value is 1.56 and the numerical
value is 1.55, however the random coin-toss limit is 1.66. We note that the
numerical results obtained using the full and reduced
strategy spaces are very similar; hence our results and conclusions
are indeed general. Finally, we would like to point out
that the high-$T$ regime has been the subject of much debate recently. In particular,  Challet
{\em et al} identified problems
\cite{comment,reply,marsili2} with the $T\rightarrow \infty$ results of Ref. 
\cite{sherrington}. We refer to Refs. \cite{comment,reply,marsili2} for a detailed
discussion. 

Given the complexity of these many-agent games, it is remarkable
that our simple analytic approach invoking crowd effects can quantitatively explain the
main feature whereby
$\sigma$ falls below the random, coin-toss limit as $\theta$ (or
$T$ \cite{sherrington}) increases.  This feature, which is arguably the most
striking result of the TMG, can therefore be explained without having to solve
the full game dynamics. This result also strengthens our belief that many properties of
the MG can be understood using simple notions of crowd-anticrowd interplay.

\newpage 
\centerline{\bf Figure Captions}

\bigskip

\noindent Figure 1: Comparison between numerical simulations (circles) and the
present analytic theory for the standard deviation 
$\sigma$ as a function of the probability $\theta$. Analytic theory (solid line):
$\sigma_{\rm delta}$ using Eq. (8), $\sigma_{\rm flat}$ using Eq. (9). $N=101$,
$m=2$ and $s=2$. Dashed line shows random
coin-toss value. Solid arrow indicates theoretical value
$\sigma_{\theta\rightarrow 0.5}=4.7$ for $\theta\rightarrow 0.5$ 
(see text). Inset shows distribution of $\sigma$ values at $\theta=0.5$ for
several thousand randomly-chosen initial strategy configurations: (a) numerical simulation,
(b) semi-analytic theory (see text). Quantities shown are dimensionless. 
\bigskip


\begin{thebibliography}{99}
\bibitem{tmg} The Thermal Minority Game discussed in Ref. \cite{sherrington}
depends on a parameter
$T$ (or equivalently $1/\beta$) called a `temperature'. We could similarly
define 
$T$ by setting the probability of playing the worst strategy
$\theta = e^{-\beta}/(e^{\beta}+e^{-\beta})$. Hence $T=2[{\rm ln} (\theta^{-1}
-1)]^{-1}$. $T=0$ corresponds to $\theta=0$ while
$T\rightarrow \infty$ corresponds to $\theta\rightarrow 1/2$, hence we will
only consider $0\leq\theta\leq 1/2$ in this paper.
\end{thebibliography}
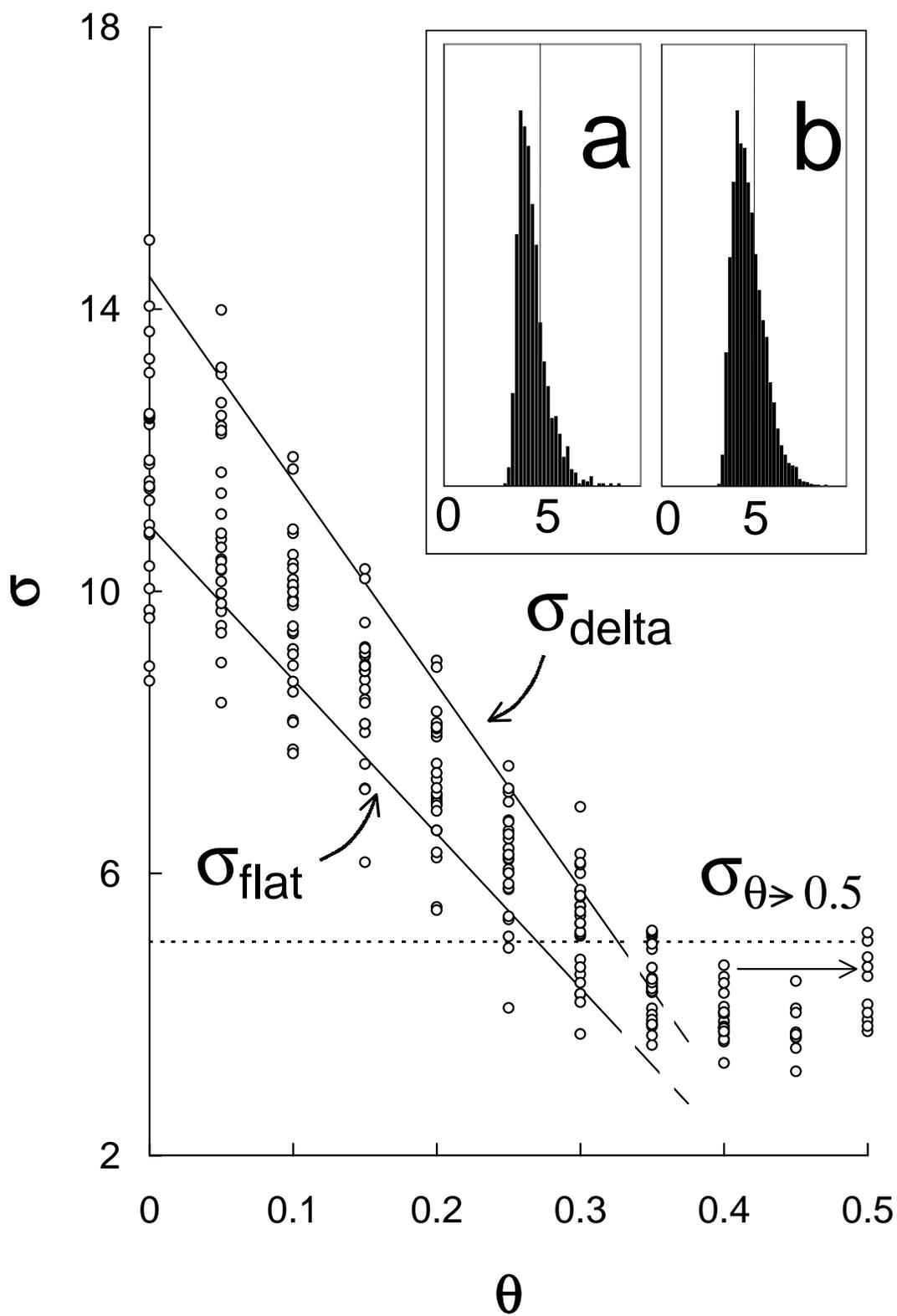
\end{document}